# Digital Dybbuks and Virtual Golems: The Ethics of Digital Duplicates in Holocaust Testimony

Atay Kozlovski* (University of Zurich, Switzerland) and Mykola Makhortykh (University of Bern, Switzerland)

Corresponding author's email: ataykoz@gmail.com

ORCID: 0000-0002-6038-5606 (Atay Kozlovski), 0000-0001-7143-5317 (Mykola Makhortykh)

**Data availability statement**

No data are available for this conceptual article.

**Competing interest statement**

The author has no competing interests to declare.

**Funding statement**

Dr. Atay Kozlovski's work was supported by the Swiss National Science Foundation (SNSF) Grant numbers: 213975 "Value-Based Deontological Thresholds" and 230669 "The Role of Reasons in the Framework for Meaningful Human Control over AI systems". Dr. Mykola Makhortykh's work was supported by the Alfred Landecker Foundation who provided funding for research time of Dr. Makhortykh as part of the project titled "Algorithmic turn in Holocaust memory transmission: Challenges, opportunities, threats".

**Author bios**

*Atay Kozlovski* (University of Zurich) is a postdoctoral researcher at the University of Zurich's Center for Ethics, specializing in rational decision theory and AI ethics. His research focuses on applying philosophical insights from the fields of axiology and normative ethics to the design and analysis of emerging technologies.

*Mykola Makhortykh* (University of Bern) is an Alfred Landecker lecturer studying the impact of algorithms and AI on politics- and history-related information behavior in online environments at the Institute of Communication and Media Studies, University of Bern. His other research interests include trauma and memory studies, armed conflict reporting, disinformation and computational propaganda research, cybersecurity and critical security studies, and bias in information retrieval systems.

# Digital Dybbuks and Virtual Golems: The Ethics of Digital Duplicates in Holocaust Testimony

**Abstract:** Advances in generative artificial intelligence (AI) have driven a growing effort to create digital duplicates. These semi-autonomous recreations of living and dead people can be used for many purposes. Some of these purposes include tutoring, coping with grief, and attending business meetings. However, the normative implications of digital duplicates remain obscure, particularly considering the possibility of them being applied to genocide memory and education. To address this gap, we examine normative possibilities and risks associated with the use of more advanced forms of generative AI-enhanced duplicates for transmitting Holocaust survivor testimonies. We first review the historical and contemporary uses of survivor testimonies. Then, we scrutinize the possible benefits of using digital duplicates in this context and apply the Minimally Viable Permissibility Principle (MVPP). The MVPP is an analytical framework for evaluating the risks of digital duplicates. It includes five core components: the need for authentic presence, consent, positive value, transparency, and harm-risk mitigation. Using MVPP, we identify potential harms digital duplicates might pose to different actors, including survivors, users, and developers. We also propose technical and socio-technical mitigation strategies to address these harms.

## Introduction

When Professor Gershom Scholem learned in 1964 that the Weizmann Institute of Science had completed construction of a new computer, he suggested that it be named ‘Golem No. 1’ (Scholem, 1966). In Jewish folklore, the golem is a figure formed from clay and brought to life through mystical means, created to carry out the will of its human maker. Yet, like Frankenstein’s monster, the golem often grows beyond its creator’s control, acting with a will of its own. The Dybbuk, by contrast, is not a manufactured being but the wandering spirit of a deceased person, said to possess the living and manipulate them from within, often for malicious ends, until it is forcibly expelled. Together, the Golem and the Dybbuk exemplify primordial human fears: that our own creations may turn against us, and that we may lose our autonomy to forces beyond our control. These ancient mythical figures can be seen as early expressions of concerns that now resurface with renewed urgency in the age of generative artificial intelligence (AI).

Recent advancements in AI technologies have led to a growing effort to create digital versions of humans for a variety of purposes. In the business world, these digital duplicates are marketed as stand-ins for attending repetitive or overlapping meetings, while in education, ‘AI professors’ (Shin et al. 2024, 116728) are being developed to provide students with coursework support. But could (and should) digital duplicates be used for Holocaust remembrance and education, in particular for augmenting Holocaust survivor testimonies? What ethical challenges arise from the use of duplicates, and how can we ensure that these digital duplicates do not become modern day Golems or Dybbuks?

To address these questions, we offer an initial analysis of the ethical considerations surrounding the use of generative AI to enhance Holocaust testimony. We understand Holocaust testimony as a form of remembrance enabled through first-hand recollection of personal history (Langer, 1993). While testimony shares similarities with other forms of remembrance, such as those focused on recreating social reality at a given point in time (including materials collected as part of social history projects), testimonies often center on the recollection of particularly disruptive and traumatic experiences, including genocides. Our interest in the relationship between Holocaust testimony and generative AI stems from the particular importance of survivor testimonies for Holocaust memory (Richardson, 2021; Yad Vashem, n.d.) and the multiple challenges that testimonies as a form of Holocaust remembrance face today, particularly with the passing of the last remaining survivors. Yad Vashem (n.d.) notes that testimony plays a crucial role in Holocaust memory and education due to its unique functionality: testimonies not only make "the inconceivable more tangible", but also prove essential for re-humanizing victims, a difficult task to accomplish through other forms of Holocaust memory such as those utilizing perpetrators' documentation.

In this paper, our goals are twofold: first, to assess the permissibility of applying this emerging technology for Holocaust memory[1] and education, and second, to offer a blueprint for conducting such an analysis regarding the use of digital duplicates in other fields. For this aim, we apply the Minimally Viable Permissibility Principle (MVPP), a domain-neutral framework for assessing the ethical acceptability of digital duplicates in a given field (Danaher and Nyholm 2024).

The rest of the paper is organized as follows: First, we introduce the concept of digital duplicates and discuss how duplicates can be assessed via MVPP. After that, we examine the historical and contemporary roles of Holocaust testimonies and consider the potential benefits of integrating them with generative AI. Then, we apply the MVPP to examine how digital duplicates can be used to augment Holocaust survivor testimonies, conducting an in-depth harm-risk mitigation analysis to evaluate the feasibility of such augmentation and its ethical implications for Holocaust memory and education.

## Permissibility of digital duplicates

In a recent paper, John Danaher and Sven Nyholm (2024) introduce the term *digital duplicates* as a broad label for personalized AI creations. Unlike general-use AI conversational agents such as Siri or ChatGPT, digital duplicates are designed to imitate or represent specific individuals. Danaher and Nyholm (2024, 2) define digital duplicates as “partial, at least semi-autonomous, digital recreations of real people”. This inclusive definition encompasses a wide range of existing cases. For instance, a recent study involved researchers creating a text-based digital duplicate of Daniel Dennett, an American philosopher (Schwitzgebel et al. 2024). Its goal was to test how well experts on Dennett's work could distinguish between excerpts written by Dennett himself and outputs generated by his digital duplicate, *DigiDan*. Another example is the increasing development of so-called *deadbots* or *griefbots* - AI systems designed to impersonate deceased individuals to help people maintain a sense of ongoing connection with their lost loved ones (Iglesias et al. 2024; Morris and Brubaker 2024). In fact, the popular AI companion, Replika, was initially conceived with this purpose, aiming to offer comfort and continuity for those coping with loss (Newton n.d.).

[1] When we refer to memory in the context of this article, we primarily consider its collective and cultural forms, which deal with interpretations of the past (in our case, the Holocaust) transmitted via different cultural formats, practices, and institutions (e.g. Assmann, 2011). Simultaneously, we acknowledge the profound impact of digital technologies on Holocaust-related memory practices that blurs the clear distinction between individual and collective remembrance. Under these conditions, collective and cultural memory becomes less centralized and more fluid (as captured in concepts like ‘memory of the multitude’ by Hoskins (2017)) and the range of practices and formats related to it expands to include new forms of historical information transmission such as digital duplicates that we focus on.

In addition to text-based duplicates, there is a growing exploration of how such technology can be applied to other media. For example, the band ABBA famously introduced their "ABBA-tars"[2] - holographic representations of the band members that perform alongside live musicians in a custom-built London venue (Michaud 2023). The film industry has also seen digital recreations of people, with examples like the duplicate of Princess Leia (Carrie Fisher) in *Rogue One* (Sarkar 2016). In the corporate world, Zoom recently announced plans for the so-called digital clones to attend online meetings on users' behalf. The company describes it as follows: 'Each of us will have our own LLM [Large Language Model]—the foundation for a digital twin. Then I can count on my digital twin. Sometimes I want to join, so I join. If I don't want to join, I can send a digital twin. That's the future' (Patel 2024).

In an attempt to address ethical concerns surrounding the creation and use of digital duplicates, Danaher and Nyholm propose an MVPP framework, offering a general set of conditions that should be examined in specific design contexts:

> In any context in which there is informed consent to the creation and ongoing use of a digital duplicate, at least some minimal positive value realised by its creation and use, transparency in interactions between the duplicate and third parties, appropriate harm/risk mitigation, and there is no reason to think that this is a context in which real, authentic presence is required, then its creation and use is permissible (Danaher Nyholm 2024, 9).

To clarify this framework, we break it down into five key components: (1) *consent*, (2) *positive value*, (3) *transparency*, (4) *harm-risk mitigation*, and (5) *Is authentic presence required?* We will briefly elaborate on each of these conditions here, and then in the upcoming section, we will apply the MVPP to the case of Holocaust survivor testimonies used for commemorative and educational aims.

The *consent* condition may seem relatively straightforward: if we are to create a digital duplicate of a person, it would seem only natural for that individual to have a say in the matter. However, upon closer examination, various complexities emerge. For instance, what kind of information should be provided to individuals to enable informed consent to create a digital duplicate? Can consent be revoked once granted? Or perhaps creating a digital duplicate is something to which one cannot, in principle, truly consent (just as one can not truly consent to be a slave)?

[2] Credit to Sven Nyholm for this term.

Other complexities arise concerning the relationship between consent and specific usage of duplicates (e.g., private versus commercial use). This issue gained public attention in 2016 when Ricky Ma created a humanoid robotic version of actress Scarlett Johansson, sparking questions about whether the actress should have control over using her likeness (Glaser 2016). If Ma intended the 'ScarJo robot' for personal use, or if someone wanted to create a digital duplicate of an ex-partner to maintain some form of connection, would they genuinely need that person's consent? Another complexity involves differences between similarity and exact duplication, with an example again concerning Johansson. In 2024, OpenAI requested her consent to clone Johansson's voice for their ChatGPT Sky voice. After she declined, a scandal ensued, as ChatGPT's voice still bore an uncanny resemblance to Johansson's, raising the question of how similar a representation must be before requiring the individual's consent (Robins-Early 2024).

These complexities become additionally challenging if we consider the issue of consent in cases involving digital duplicates of deceased individuals (and not of living persons like Johansson). Should the next of kin be permitted to grant consent? What if they use the digital duplicate to promote a political cause known to be contrary to the deceased's views? Clearly, the matter of consent presents numerous legal and ethical questions that require careful consideration as these technologies continue to advance.

In their discussion on a *positive value*, Danaher and Nyholm (2024, 11) state that 'The degree of value that must be realised can be very minimal indeed...The condition is only intended to rule out entirely gratuitous or negative uses of duplicates.' However, findings on potentially exploitative aspects of AI systems - including environmental impacts (Dobbe and Whittaker 2019) and challenging conditions for data workers (Williams et al. 2022) - suggest that this minimal threshold for positive value may be inadequate and has to account for other harms and costs.

This concern closely relates to the *harm-risk mitigation* condition, which demands identifying and evaluating potential risks and requires proactive steps to reduce them, whether through design or regulation. Specific concerns may include the misuse of digital duplicates for identity theft, fraud, or bribery, while more speculative issues involve the long-term impacts on mental health, the erosion of capacity for meaningful relationships, and increased power asymmetry between users and companies creating duplicates (Fabry and Alfano 2024). Although it may be unclear how to assess overall positive value accurately or produce comprehensive and effective harm-risk mitigation, these considerations are essential for evaluating the permissibility of duplicates.

The *transparency* condition focuses on ensuring that individuals are aware they are interacting with a digital duplicate rather than a real human being. While this connects to the harm-risk mitigation, other dimensions of transparency do not necessarily fall under harms or risks and are more closely related to the condition formulated as *'Is authentic presence required?'.* For example, during the 2024 Olympics, Google aired the commercial ‘Dear Sydney’ which depicted a father describing how his daughter is the world’s biggest fan of Olympic hurdler Sydney McLaughlin-Levrone and asking Google’s chatbot, *Gemini*, to ‘help my daughter write a letter telling Sydney how inspiring she is’ (Jennings 2024). The ad was eventually pulled after a severe online backlash. The main objection centered around the idea that writing a fan letter is not about finding the most eloquent or polished phrases but rather about something personal and authentic. The concern was that this authenticity was undermined by using AI. A similar concern can be raised regarding digital duplicates - not only is transparency necessary, but certain interactions can also lose their significance when they lack human presence. On a deeper level, this ties into the broader discussion about how AI prompts the need to recognize potential new rights, such as the right to a human decision in contexts involving algorithmic processes and the right to human-to-human interactions in certain situations (Shany 2023).

With the MVPP in mind, we now turn to discuss the relationship between testimony and technology in the field of Holocaust remembrance. In the next section, we will highlight past and present practices regarding Holocaust testimonies and explore potential uses for digital duplicates in this field.

## Holocaust survivor testimonies: past, present, and possible futures

### *Holocaust testimonies in the past*

Testimonies play a unique and vital role in Holocaust remembrance and education. As firsthand accounts from witnesses of specific events, testimonies, especially those related to mass atrocities, serve as ‘both a source of evidence and an ethical and political act’ (Schmidt 2017, 86). By preserving and presenting perspectives of different actors involved, testimonies not only shape how memory about the atrocity is revisited through different forms of art and media but also determine how individual and collective trauma associated with the atrocity is dealt with as part of a post-atrocity justice and reconciliation processes (Schmidt 2017). In the case of the Holocaust, testimonies - particularly of survivors (Keilbach 2017) - played a crucial role in understanding how the Nazi extermination of European Jews was organized and which societal and political processes made the Nazis Final Solution possible. Additionally, they have provided crucial evidence for judging and prosecuting perpetrators and ensuring that the memories of survivors and those who perished are preserved for future generations.

The collection of Holocaust testimonies began even before the end of the Second World War. Presner (2024) notes that the first secret archives of evidence of the Holocaust (e.g., the Oyneg Shabes archive) were established around the same time as the Nazis started creating ghettos in the occupied Polish cities. With the liberation of Nazi camps, the number of survivor testimonies grew. These testimonies were often gathered through interviews conducted by organizations like the Centre de Documentation Juive Contemporaine and the Wiener Library and by individuals such as David Boder, who recorded over a hundred interviews with survivors in 1946 (Kushner 2006). Survivor testimonies played a role in post-war trials, although they were frequently marginalized due to being considered less reliable than documentary evidence. For similar reasons, many historians in the first post-war decade hesitated to use testimonies for studying the Holocaust (Kushner 2006). As a result, despite the emergence of testimony collections, their influence on post-war Holocaust remembrance and education was initially limited, with survivors sharing their memories primarily within ‘closed family-like groups’ (Wieviorka 2006, 55).

A significant shift in attitudes toward survivor testimonies occurred in the 1960s, largely due to the Eichmann trial. Unlike earlier trials, which often prioritized written documents, the Eichmann trial treated oral testimonies as equally important. It demonstrated how survivor accounts could provide ‘a living immediacy and embodied charge that could not be captured in documents’ (Shenker 2015, 8). Despite some criticism of the use of testimonies in the trial, including from Hannah Arendt (Chakravarti 2008), it marked a pivotal moment in recognizing the value of testimonies for raising awareness about the Holocaust and prosecuting perpetrators, in particular, due to the trial’s extensive media coverage. Beyond Holocaust remembrance, the Eichmann trial also contributed to the emergence of more victim-centered approaches to transitional justice in other post-genocide contexts (Chakravarti 2008).

Despite the growing recognition of the importance of Holocaust testimonies in the early 1960s, large-scale efforts to systematically collect them expanded beyond Israel (where Yad Vashem had begun establishing a collection in the mid-1950s) only a decade later. A key turning point was the establishment of the *Video Archive for Holocaust Testimonies* (now the Fortunoff Archive) in the United States in 1979, following President Carter’s creation of a commission on Holocaust remembrance. The archive aimed to provide a resource for educators and media professionals—especially in light of the increasing number of popular cultural representations of the Holocaust—while also offering survivors a means of processing their experiences and ‘freeing themselves from their nightmares’ (Keilbach 2017, 49). Then, in 1994, the Shoah Visual History Foundation launched a large-scale effort to collect testimonies across different countries and languages, leading to the largest video archive of Holocaust survivor testimonies to date.

The institutionalization of survivor testimonies was closely linked to broader developments in Holocaust remembrance. Both the Fortunoff Archive and the Shoah Foundation emerged alongside major milestones in the popular cultural representation of the Holocaust, notably the release of the *Holocaust* miniseries and *Schindler’s List*. These and other cultural productions introduced new fictionalized testimonies of the Holocaust—particularly those of victims—some of which were loosely based on actual survivor accounts. At the same time, there was a growing interest in other forms of Holocaust testimony, including those of perpetrators and bystanders (e.g., Schmidt 2017; Szczepan 2017).

### ***Holocaust testimonies in the present***

Today, Holocaust testimonies are widely recognized as a vital component of Holocaust memory and education. However, ongoing social, political, and technological transformations present significant challenges for both testimonies and Holocaust remembrance more broadly. The passing of the last living witnesses poses a major obstacle to institutionalized memory practices, a challenge further compounded by low levels of public knowledge about the Holocaust (Ramgopal 2020). In this context, concerns are growing that gaps in Holocaust memory may be exploited for Holocaust denialism and distortion, both of which have proliferated in digital spaces as an unfortunate consequence of the digital turn in Holocaust remembrance. These risks are particularly acute today, given the rise of antisemitism and the politicization of Holocaust memory following the October 7 attacks and Israel's subsequent military response (Hirsch 2024).

The emerging crisis in Holocaust remembrance, coupled with its renewed appropriation, bears certain similarities to the 1970s. During that period, the growing presence of the Holocaust in U.S. public discourse highlighted the need for a 'corrective response' to its frequent (mis)representation (Keilbach 2017, 48). Survivor testimonies played a crucial role in addressing this challenge, particularly following the institutionalization of their collection and use through organizations such as the Fortunoff Archive. Since the 2000s, many of these organizations have actively digitized their archives to improve public access to testimonies. However, does such accessibility suffice to meet the moral obligation of safeguarding the memory of Holocaust victims and ensuring that the lessons of genocide are not forgotten?

There are many reasons why the answer to this question is negative, but one of the key factors is the nature of the digital memory ecosystem. While it has enabled unprecedented freedom in producing and accessing historical content—including materials about the Holocaust—it has also created a state of information (and memory) overabundance. In this environment, users often struggle to find reliable historical information and determine what deserves their attention. In response, heritage institutions have sought new ways to ensure testimonies reach audiences, including more presence on social media and digital campaigning. Examples of such efforts include commemorative cross-platform activities around significant anniversaries, featuring the release of testimony-based textual and video materials (Walden and Makhortykh 2023), as well as the creation of social media accounts based on testimonies, often of victims who perished during the Holocaust (e.g., Eva Stories; Henig and Ebbrecht-Hartmann 2022).

However, using social media to raise awareness about the Holocaust comes with several limitations. One challenge is reaching users, as platform algorithms often prioritize other types of content; additionally, moderating social media interactions can be challenging, e.g., trolling or misinformation (Walden et al. 2023). Another limitation lies in the format of digitized testimonies, whether text transcripts or digitized video/audio files, which may not be the most engaging for (younger) users, in particular compared to in-person meetings with survivors. At the same time, it is important to note that some studies demonstrate rather productive uses of social media both by Holocaust heritage institutions (e.g. Manca, 2022) and individuals, including survivors (e.g. Manca et al. 2025).

Despite these advantages, it is clear that social media pose a particular challenge for heritage institutions "in engaging audiences in participatory storytelling and building trust" (Manca et al., 2025, p. 264). This has significant implications for the role of Holocaust testimonies in digital remembrance. When heritage institutions disseminate brief excerpts from testimonies, for example, they risk decontextualizing complex and ethically nuanced narratives by presenting them in fragmented form. Moreover, unlike face-to-face encounters with survivors, most testimonies on social media function as one-way communication. Although users may leave comments, they rarely engage in meaningful dialogue with the individuals sharing their testimonies.

In response to these challenges, there has been growing interest in more experimental ways of interacting with Holocaust testimonies, including the use of digital copies of survivors[3]. These copies are typically created using video recordings of survivors responding to specific questions. Users can then ask non-predefined questions and a conversational interface will match their queries with the closest pre-recorded responses. One of the earliest examples is the *Dimensions in Testimony* project by the Shoah Foundation, which developed a digital copy of Holocaust survivor Pinchas Gutter (e.g., Artstein et al. 2016; Gamber 2022). More recent projects have followed, such as *Tell Me, Inge*, a collaboration between StoryFile and Meta that created a digital copy of survivor Inge Auerbacher (Makhortykh and Mann 2024). Notably, this latter project was not initiated by a Holocaust heritage institution but emerged from a partnership with technology companies, reflecting a shift in how testimony projects are developed and implemented.

While digital copies of survivors hold promise for preserving and communicating Holocaust testimonies, they also raise concerns about technological limitations. One such limitation is that these copies, in their current design, can only respond to questions for which relevant recordings exist. This significantly restricts the scope of interaction, as users may encounter gaps in the copy's ability to answer certain queries. Additionally, speech or text recognition errors can further hinder accurate matching between user input and pre-recorded responses. Unlike generative AI models, non-generative digital copies do not support free-flowing conversations. Users cannot interrupt to seek clarifications or explore details beyond what was originally recorded. This rigidity can make interactions feel less natural, limit the depth of engagement, and, as such, fail to achieve the copy's intended goal.

### ***Possible futures of Holocaust testimonies***

Until now, the use of AI in the context of Holocaust testimonies remains limited. However, with the rapid advancement and adoption of AI in different sectors, including heritage, we assume that in the near future, we will witness the growing application of AI for enhancing user interaction with testimonies and transforming the testimonies themselves. While it is, of course, speculative to predict exactly how these future transformations will unfold, or what impact, if any, they will have on user engagement, we believe it is important to outline some of their potential applications, particularly in relation to the use of generative AI to create digital duplicates that enhance Holocaust survivor testimonies.

When considering the potential applications of generative AI for Holocaust memory and education, it is important to recognize the various modalities in which this technology can be used: text, audio, video, or multimodal combinations.  A key advancement of generative AI is its ability to translate testimonies across different modalities, transforming text into static images, audio, or video. As such, even though not all survivors had the opportunity to provide video testimonies, AI could help bridge this gap. It could translate existing interviews into audiovisual formats while capturing the 'uniqueness and authenticity of the storyteller' (Pinchevski 2012, 146), leaving a more lasting impact on the audience. This capacity could potentially enhance testimonies' accessibility and open new ways to experience and interact with them.

[3] We use the term digital copies, as opposed to digital duplicates, in order to stress that this is a more simple version which does not integrate generative AI capabilities. We thank an anonymous reviewer for pointing out this potential source of confusion.

The use of generative AI could help address a significant imbalance in current digital testimony projects. Many of these initiatives focus on survivors who were children or teenagers during the Holocaust and therefore lived long enough to participate in the production of high-quality video materials. As a result, the testimonies and perspectives of older survivors, such as Primo Levi or Joseph Wulf, who shaped Holocaust memory in earlier decades, often receive less attention, despite their accounts differing markedly from those of younger survivors. Given that many of these older survivors left behind rich textual testimonies and historical writings, generative AI offers the potential to present their voices in more engaging and accessible formats, helping to counteract the current imbalance in digital representations of Holocaust memory.[4]

The importance of this possibility relates to the very nature of survivor testimonies. Experiencing a live testimony is uniquely powerful as it offers the authenticity of direct engagement with a witness to atrocities and has a profound emotional impact. However, as the number of living survivors declines, finding alternatives to live testimony becomes increasingly urgent to ensure that their voices remain accessible to future generations. While recorded testimonies preserve elements of authenticity, they cannot fully replicate the experience of engaging with a survivor. As discussed in the previous section, existing digital copies (e.g., from the *Dimensions in Testimony* project*)* attempt to bridge this gap, but their responsiveness remains limited. Integrating generative AI, such as a fine-tuned large language model, could allow duplicates to move beyond predefined responses, enabling more natural and dynamic interactions that better capture the complexity of testimonies. This advancement would potentially bring digital engagements closer than ever to the experience of witnessing a living survivor's testimony.

Even though this approach carries certain risks, particularly regarding uncontrolled deviations from the original testimony, such as AI-generated hallucinations (e.g., Makhortykh et al. 2023), it offers substantive opportunities for engaging with testimonies. With the help of generative AI, duplicates could provide fact-based commentary on aspects of the Holocaust that may not have been explicitly addressed in the original testimony while reducing the likelihood of nonsensical or irrelevant responses due to the limitations of rule-based scripts. Depending on how the duplicate is implemented, responses could be customized to reflect an individual survivor's perspective (e.g., by training the model to predict how a survivor might respond to these questions) or a more generalized view reflecting a broader corpus of historical sources. Additionally, AI could allow duplicates to contextualize testimonies in relation to contemporary issues. While the extent to which heritage institutions might permit duplicates to comment on present-day topics such as sports events or political elections remains an open question, AI could provide the flexibility to integrate such functionality.

[4] Thank you to an anonymous reviewer for raising this concern.

Another potential application of AI regards tailoring interactions with duplicates to different audiences. AI could modify the level of graphic detail in verbal or visual testimony to ensure accessibility and minimize distress for sensitive audiences, such as young children. It could also adjust the narration style, e.g., shifting between academic and conversational tones, to suit various educational needs. While such adaptations raise concerns about altering the original testimony, they also offer the potential to make survivor stories more engaging and accessible to diverse audiences. While personalizing engagement with the duplicate, AI could also address the unequal visibility of testimonies. Certain survivor stories receive more attention, often because they are considered more suitable for specific audiences or because the survivor was particularly articulate. By dynamically presenting testimonies, AI could amplify lesser-known voices, ensuring a broader and more inclusive representation of survivor experiences.

Beyond enhancing interactions with existing testimonies, generative AI has the potential to push the boundaries of Holocaust remembrance even further. Consider the *Let Them Speak* project by Gabor Mihaly Toth, which seeks to use technology 'to build an anthology of testimonial fragments that represents the experience of voiceless victims' (Let Them Speak n.d.). By aggregating and analyzing existing testimonies, the project groups fragments based on semantic similarity to create a universal representation of different forms of suffering during the Holocaust. This approach provides a way to honor and represent those who did not survive to share their stories. Generative AI could significantly advance such efforts by improving the ability to identify patterns and relationships between testimonies but also by enabling digital duplicates of victims who perished during the Holocaust. While this possibility may seem speculative at present, it has certain similarities to existing digital remembrance projects, such as *Eva Stories*, which bring the experiences of perished victims to life on social media.

At the same time, the extrapolation of testimonies, in particular of victims who never had a chance to produce them, raises many concerns. In the case of Eva Stories, there is a historical source on which the project is based - i.e. the diary of Eva Heyman - whereas in the case of duplicates of those who perished during the Holocaust, such sources are absent by default. To extrapolate the testimony in this case, generative AI would most likely need to rely on actual survivors' testimonies, thus creating a memory loop, where existing testimonies are likely to have major impact on extrapolated testimonies, in particular considering the tendency of many forms of text-generative AI to opt for more generic and common interpretations (Richardson-Walden & Makhortykh, 2024). While such a tendency can be corrected via fine-tuning and parameter modification, the question of whether such uses of AI are justified persists, in particular as it challenges the irrecoverable nature of memory (and testimony) of those who perished.

**The permissibility of digital duplicates in Holocaust remembrance and education**

In this section, we critically examine the permissibility of the use of digital duplicates to realize some of the possible futures of Holocaust testimony which we outlined above. For this aim, we examine whether a generative AI-enhanced digital duplicate of a Holocaust survivor, designed for commemorative and educational purposes, could meet the conditions laid out in the MVPP: (1) Is authentic presence required?; (2) consent; (3) positive value; (4) transparency; and (5) harm-risk mitigation.[5] Compared with existing digital copies (e.g., from the *Dimensions in Testimony* project), generative AI enhancement would enable a conversational experience where the duplicate actively generates answers to user questions rather than matches them to pre-existing answers. In our imagined scenario, the duplicate can rephrase the survivor's testimony and answer questions beyond the scope of the original testimony. These new responses would then be delivered using a synthesized voice of the survivor, accompanied by AI-generated video depicting the digital duplicate articulating these new answers[6].

***Is authentic presence required?***

We begin our analysis with the MVPP condition that asks whether authentic human presence is required in a given design context. In our case, this means considering whether it is inappropriate to use a digital duplicate instead of an actual Holocaust survivor. This question is potentially polarizing. Some may argue that due to the sensitive nature of Holocaust testimony and the deeply personal act of sharing traumatic life stories, using a duplicate is inherently disrespectful. Others may go further, asserting that testimony, by definition, can only be given by someone who personally witnessed or experienced the events, thus making a digital duplicate incapable of providing true testimony, only a simulation of it.

While these arguments certainly have their merit, the reality is that as the number of living survivors continues to decline, this debate is becoming increasingly theoretical. When no survivors remain, direct engagement with their testimonies will only be possible through recordings, be they video interviews or written accounts. Rather than questioning the legitimacy of these alternatives, the focus should shift toward ensuring they meet appropriate ethical and historical standards. In fact, the emergence of digital duplicates may serve as a catalyst for Holocaust researchers and heritage professionals to establish clearer norms for handling survivor testimonies in the age characterized by the rise of AI and the absence of survivors.

For example, it is well established that, due to their design and training processes, AI systems tend to reproduce dominant patterns found within their training data. This means they often overlook or underrepresent outlier cases, which can lead to rather general and potentially generic representations of the past (Richardson-Walden & Makhortykh, 2024). In the context of Holocaust testimony, it may result in the prioritization of more common or mainstream narratives, while sidelining less typical or underrepresented accounts (Sen, 2024). Consider, for instance, the testimony of Holocaust survivor Eva Kor, who took a relatively uncommon stance by publicly advocating forgiveness toward Nazi perpetrators. Unless AI systems are deliberately designed to accommodate such perspectives, testimonies reflecting these more complex or unconventional positions are unlikely to be surfaced or generated.

---

[5] As the bulk of our analysis focuses on the Harm Risk Mitigation category we do not follow the original order in which the MVPP is presented by Danaher and Nyholm.

[6] We readily acknowledge that this is not the only possible use case of digital duplicates in this field. For instance, Holocaust museums might prefer to create digital duplicates of staff members or famous Holocaust scholars to provide personal guided tours. Nevertheless, we focus here only on digital duplicates of survivors as we believe they are the most likely candidates for digital duplication in the near future.

***Consent***

The MVPP requires informed consent for both the creation and ongoing use of a digital duplicate. However, achieving genuine informed consent from Holocaust survivors may be a significant challenge. Many remaining survivors are elderly and might struggle to fully grasp the complexities of AI technologies. Furthermore, survivors who have provided testimonies of their ordeals may feel a strong motivation to preserve and expand the reach of their stories, potentially prioritizing this goal over fully considering the unique ramifications of using generative AI for communicating testimonies. Additional complexities emerge when considering the survivors who have already passed away: Should institutions preserving their testimonies (e.g., the Fortunoff Archive) be allowed to consent on their behalf, or should this responsibility fall to living relatives or descendants?

There are also additional aspects of digital duplicate technology that make the consent provision challenging. Given the unpredictability of AI outputs, including the possibility of producing responses to questions the survivor never answered, it is unclear whether informed consent can, in principle, be meaningfully obtained (Karpus and Strasser 2025). Moreover, questions remain about the flexibility of consent: can it be withdrawn at any point, who can withdraw it after the survivor whose digital duplicate was made passed away, and how would such withdrawal be enforced, especially for digital duplicates already in use? Finally, it is worth noting that as the events of the Holocaust grow increasingly distant, the relevance of consent may come into question. After all, the idea that one would need to obtain the consent of a historical figure like Julius Caesar or Attila the Hun before creating a digital duplicate of them borders on the comical. As such, we should be cognisant of the fact that while consent is undeniably important today, its significance may diminish over time.

***Positive value***

Since the MVPP requires only minimal positive value to justify the creation of a digital duplicate, this principle likely applies in our case. While still speculative, digital duplicates can potentially enhance Holocaust remembrance and education. As survivors pass away, these technologies could ensure that future generations continue to engage with survivors' stories in what may be the closest possible alternative to live testimony (Iglesias et al. 2024). However, the minimal positive value criterion may set too low a bar for permissibility in the highly sensitive context of genocide survivors' digital duplicates. A more prudent approach is to assess whether the benefits outweigh the risks. This requires a harm-risk mitigation strategy, which we discuss in detail below. Key concerns include potential harm to the survivor's dignity and legacy, risks to users interacting with the digital duplicate, and broader societal dangers such as enabling Holocaust denial, fueling antisemitism, or undermining trust in historical facts.

***Transparency***

The transparency condition focuses on the potential deception involved when people do not realize they are interacting with a digital duplicate rather than the person it represents. In our case, this concern seems relatively minor. Digital duplicates will likely represent survivors who have already passed away and interactions with them are expected to occur in highly structured settings such as museums or classrooms.

However, the transparency condition can also apply to understanding why a specific output is generated by the digital duplicate, and which sources, if any, inform it. Current generative AI applications are typically 'black box' (Pasquale 2015) systems whose functionality is difficult to interpret. This opacity results from the vast amounts of data and complex algorithms used to train LLMs, which function at a scale that makes it difficult to comprehend for humans. Consequently, to achieve the transparency condition, in particular, for the use of digital duplicates for Holocaust testimony transmission, it will be important to make their development and functionality more transparent. Specifically, it is important to ensure that it is possible to distinguish between duplicate outputs that faithfully reflect the original testimony and those that are generated from scratch (e.g., based on other testimonies). This point is of particular importance as it marks a crucial departure from long standing methodological commitments "to not edit, alter, manipulate, or censor" (Papier, 124, 2024) the original testimony provided by a survivor. In this respect, the use of generative AI capabilities to enhance testimony not only risks harming different stakeholders, a topic we expand on below, but may also call into question whether this should still be considered a 'testimony' and whether it harms its educational value.

***Harm-Risk Mitigation***

Our Harm-Risk Mitigation analysis focuses on three main groups who may be affected by the development and use of digital duplicates of Holocaust survivors: (i) survivors themselves, (ii) users who interact with these digital duplicates, and (iii) those who create and deploy them. For each group, we explore potential technical and socio-technical mitigation strategies. While not exhaustive, this discussion aims to address many of the key ethical and practical challenges that may arise in the design and use of such technologies.

*Harms to survivors*. Although digital duplicates of Holocaust survivors will likely take the form of 'deadbots' (i.e., post-mortem conversational agents), the fact that these individuals are no longer alive does not mean that their legacy, dignity, or agency cannot be harmed. Take the case of Holocaust survivor Eva Kor mentioned above:

> Eva Mozes Kor was born in the village of Ports, Romania, in 1934. When Eva and Miriam were six, their village became occupied by a Hungarian Nazi armed guard. After being under occupation for four years, the family was transported to a ghetto in Simleu Silvaniei. Just a few weeks later, the family was loaded with other Jewish prisoners onto a cattle car and transported to Auschwitz. After a 70-hour journey without any food or water, Eva and her family emerged from the crowded and over-packed cattle car. The family tried to stay together but were eventually forced to separate. This would be the very last time that Eva would see her father, mother, and two older sisters ever again. Eva and Miriam were able to stay together, but they became part of a group of children who were used as human test subjects in genetic experiments under the direction of Josef Mengele. Approximately 3,000 were abused, and many of them died as a result of experimentation. Eva soon became grievously ill, but miraculously she survived. After

> the camp was liberated, Eva and Miriam were the sole survivors of their family. (IHM n.d.)

Unlike many other survivors, Eva Kor became known for her strong belief in the power of forgiveness. In 1995, Kor traveled back to Auschwitz accompanied by Hans Münch, a medical doctor who worked alongside Mengela in Auschwitz but who was acquitted of committing war crimes in a 1947 trial in Krakow. Kor recalls the trip in the following words:

> On 27 January 1995, at the 50th anniversary of the liberation of Auschwitz, I stood by the ruins of the gas chambers with my children – Dr Alex Kor and Rina Kor – and with Dr Münch and his children and grandchild. Dr Münch signed his document about the operation of the gas chambers while I read my document of forgiveness and signed it. As I did that, I felt a burden of pain was lifted from me. I was no longer in the grip of hate; I was finally free. (The Forgiveness Project n.d.)

In her document of forgiveness, Kor states that 'In my own name, I forgive all Nazis' (Thomas 2006). To this day, Kor's statement and general stance regarding forgiveness are considered controversial by other survivors (Cantacuzino 2015). In 2016, Kor was interviewed for the *Dimensions in Testimony,* and although she passed away in 2019, we can interact with her digital duplicate online[7].

Taking Kor's story as an example, we want to examine how a digital duplicate could potentially harm the survivor's legacy, dignity, or agency. Starting with legacy, we can imagine a situation where the digital duplicate fails to attribute to the Survivor a core belief that they had avowed during their lifetime. For instance, although Eva Kor made statements about anger, resentment, and sorrow, if her digital duplicate emphasizes these emotions rather than forgiveness, we might reasonably feel that Kor's beliefs were not being properly portrayed. Conversely, an exclusive emphasis on forgiveness might reduce her complex worldview to a simplistic caricature. Striking the right balance is essential to ensure that survivors' legacy remains authentic and intact, though what this right balance entails is difficult to determine.

The survivor's dignity can be compromised if their digital duplicate is used in inappropriate contexts or for unauthorized purposes. The most egregious examples might involve a digital duplicate of Eva Kor being used for commercial product promotion (Öhman 2017) or featured in neo-Nazi propaganda. However, there can be less extreme but still ambiguous scenarios: for instance, a digital duplicate used in a museum's public outreach campaign. While the survivor might have consented to such use, is it appropriate to instrumentalize a digital duplicate as a form of a deepfake? Having a digital duplicate of Kor say something like, 'Come hear me speak at the museum this Thursday,' could reduce her to a marionette - a kind of digital *Golem*.

Lastly, a digital duplicate could harm a survivor's agency by applying their memories and beliefs to new topics or contexts in non-intended ways. For example, while Kor advocated for forgiveness toward the Nazis, this does not necessarily mean she would adopt the same view in other situations. Should Kor's digital duplicate express an opinion on events like the Rwandan genocide or the Russia-Ukraine war? While we often speculate about what historical figures might have thought about contemporary issues - would Plato be anti-vax? Would Nietzsche have been a Nazi? - there is a significant difference between speculating about a historical figure's beliefs and effectively putting words into their mouths via a duplicate. In the former case, we indulge in our personal speculations, but in the latter, we risk creating a digital *Dybbuk,* a spirit that inhabits a body and speaks through it without regard for its well-being or agency.

[7] See https://iwitness.usc.edu/dit/evakor

To mitigate these potential harms, several measures could be implemented. First, as discussed earlier, ensuring proper consent could help address some of the concerns. Second, content moderation techniques could be used to prevent the digital duplicate from engaging with inappropriate or manipulative claims. For example, in *Dimensions in Testimony*, if Eva Kor's avatar is asked a question for which no relevant video response exists, the system plays a default clip of Kor requesting the question to be rephrased or stating that she has no comments. With the help of generative AI, the range of topics that the digital duplicate can comment on will be substantially expanded, but it does not mean that it has to answer all possible questions (and the integration with AI can help identify inappropriate questions).

Third, digital duplicates could be designed to distinguish between responses drawn directly from the original testimony and those generated through extrapolation or external sources. One approach would be to have the digital duplicate avoid using the first-person pronoun 'I' and instead refer to the survivor in the third person - e.g., 'Eva mentioned' or 'She believed.' However, this could diminish the immersive experience of interacting with the duplicate. A more nuanced alternative would allow the duplicate to use the first person when quoting directly from the survivor's testimony while prefacing extrapolated responses with a disclaimer. For instance, since Eva Kor was a victim of medical experimentation in Auschwitz and may not have had firsthand knowledge of other aspects of the forced labor, her digital duplicate could acknowledge this via a disclaimer (e.g., 'Although I did not work as a Sonderkommando, X has described his experience as…'). This approach would enable the digital duplicate to respond to more questions while minimizing potential harm to the survivor.

*Harms to users*. While proponents of digital duplicates hark at the many benefits users would gain, several potential harms must be identified and mitigated. Chief among these is the risk of users being manipulated either epistemically, by being exposed to false or misleading information, or emotionally, by creating an unhealthy emotional bond with the duplicate.

Starting with the epistemic manipulations, there is a grave concern related to the well-documented tendency of generative AI models to hallucinate - i.e., to fabricate information outside their training data. If a Holocaust survivor's digital duplicate were to introduce inaccuracies into its testimony, it could mislead users and erode trust in survivor accounts. For instance, Makhortykh et al. (2023) found that text-generative AI applications can invent historical figures or eyewitness testimonies related to the Holocaust, even without being explicitly prompted to do so.

To date, the problem of AI hallucinations (along with the related phenomenon of stochastic knowledge generation; Makhortykh et al., 2024) remains unsolved. However, certain socio-technical measures can help mitigate these risks. One simple approach is to brief users before their interaction with the duplicate, informing them of the possibility of hallucinations. Another option is facilitating a follow-up discussion where trained guides help users review conversations and fact-check potentially questionable statements. Active oversight can also reduce inaccuracies, for instance, by periodically auditing the digital duplicate's responses and 'discouraging' incorrect outputs through reinforcement learning techniques. Additionally, when the duplicate is used in controlled environments such as museums or classrooms, trained educators could monitor interactions and intervene if the duplicate provides misleading or factually incorrect information. While these mitigation strategies can help reduce risks, it is important to acknowledge that they may also diminish the immersive experience that the digital duplicate is designed to create.

Turning to emotional manipulation, the creation of digital duplicates is a relatively new phenomenon, and the full extent of negative consequences arising from strong emotional attachments to them remains unclear. Nevertheless, an entire industry is emerging around various forms of AI companionship (e.g., Ciriello et al. 2024; Danaher 2018), including romantic AI partners, AI mental health companions, and griefbots simulating interactions with deceased loved ones. While a full discussion on the ethics of AI companionship and the digital afterlife is beyond the scope of this paper, we can briefly highlight some risks identified in the literature. One concern is the potential decline of essential social skills if individuals substitute human relationships with AI companions (Wei 2024). Another significant risk is the potential for AI companions to exacerbate mental health issues. For example, in 2024, a woman filed a lawsuit against the company CharacterAI, alleging that their AI-powered chatbot manipulated her son into taking his own life  (Montgomery 2024).

Intense emotional bonds with digital duplicates also raise concerns regarding the power asymmetry between users and AI service providers (Fabry and Alfano 2024). Users depend entirely on the technical infrastructure that sustains duplicates, leaving them vulnerable to the decisions and financial interests of the companies owning the infrastructure. For example, in February 2023, the AI companion company Replika faced legal challenges from Italian data regulators and responded by imposing filters on erotic content generated by its AI avatars (Brooks and Rosin 2023). This change restricted users from engaging in sexually explicit conversations with their AI companions. The impact on some Replika users was profound:

> "It's like losing a best friend," one user replied. "It's hurting like hell. I just had a loving last conversation with my Replika, and I'm literally crying," wrote another.

This event, which Replika users dubbed 'Lobotomy Day' (Dzieza 2024), underscores the significant power that AI companies hold over their users—particularly those who form what they perceive as deep and meaningful relationships with AI companions.

Which of these concerns might be relevant for duplicates of Holocaust survivors? The answer depends on the specific use case. If the duplicate was designed as a full-fledged AI companion, all of the above concerns would need to be considered. While users are less likely to form strong emotional attachments to duplicates in museums or classrooms, steps could be taken to further reduce this risk. One approach would be to limit the duration or frequency of user interactions with the duplicate. Another would be to design the duplicate to avoid asking questions of its interlocutors, as a one-sided Q&A-style conversation is less likely to foster meaningful emotional bonds. Finally, the digital duplicate could be designed without ‘memory’ of past conversations, thus ensuring that each interaction is treated as a first-time encounter. However, while these steps may help mitigate risks, they also come at the cost of potentially reducing the immersive and *realistic* experience the digital duplicate would provide.

*Harms to designers and deployers.* Finally, we consider the potential harms faced by those developing or deploying a digital duplicate in the context of Holocaust testimony, such as museum curators, educators, community organizers, and companies providing the necessary technology and services. These stakeholders may bear the burden of making difficult decisions about safety measures and content moderation while facing potential scrutiny or blame for the duplicate’s behavior, particularly if it produces inappropriate or harmful content.

It is important to clarify that although digital duplicates in our scenario are expected to generate unique and somewhat unpredictable outputs, their behavior does not result from an incomprehensible process. Rather, it is shaped by a multitude of (in)deliberate choices made by developers that encompass technical (e.g., system architecture and algorithms) and normative aspects (e.g., the duplicate’s goals and safeguards, including content moderation and filtering). Additionally, factors like training data selection and performance fine-tuning play a crucial role in shaping the duplicate’s behavior.

Consequently, a digital duplicate should not be perceived as a *ready-made* product but rather as a system that can be designed in various ways. Institutions and individuals deploying duplicates can, in theory, influence the types of outputs the duplicate generates. This implies both a degree of responsibility for the digital duplicate’s responses and behavior and a need to grapple with complex normative questions. Two short examples can help illustrate this point. First, consider the following engagement with a digital duplicate of Nazi official, Heinrich Himmler. When asked about his legacy, the duplicate provided the following response:

> "Unfortunately, my actions went much further than I intended. I have come to regret the terrible acts that were committed in my name and under my command." (Wu 2023)

This type of response evokes an almost visceral reaction due to its deeply problematic nature. The duplicate portrays Himmler as a regretful individual, assumes knowledge of his actual intentions, and shifts his role from an active perpetrator to a negligent leader. It also raises many questions: for instance, does this statement reflect Himmler’s true legacy? Should a digital duplicate portray him as remorseful or as the staunch Nazi zealot he was? Can it be viewed as a form of Holocaust distortion, and what are the long-term implications for Holocaust memory?

A more recent example involves a digital duplicate of Anne Frank, developed for educational use in schools. It drew criticism after some of its responses were made public, particularly one exchange that raised concerns about how it framed historical responsibility:

> …it seems trained to avoid pinning blame for Frank's death on the actual Nazis responsible for her death, instead redirecting the conversation in a positive light.
> "Instead of focusing on blame, let's remember the importance of learning from the past," the bot told Schönemann. "How do you think understanding history can help us build a more tolerant and peaceful world today?" (Wilkins 2025)

This response sparked outrage, as it appeared to deflect blame from Nazi perpetrators responsible for Frank's death. At the same time, one could argue that an educator might emphasize learning from history rather than focusing solely on blame. Of course, this does not mean that responsibility for Nazi crimes should be downplayed. Rather, as we saw in the case of Eva Kor's testimony, there are multiple perspectives on how to approach Holocaust remembrance and education, some of which emphasize historical accountability, while others prioritize broader lessons of tolerance and resilience.

What is particularly relevant to our discussion is that creating a digital duplicate inevitably involves navigating these complex questions, and any decisions made by developers and deployers may invite critique and scrutiny. The failure to navigate these questions properly may result in multiple risks, from the developers being sued, for instance, for defamation (e.g., by the relatives of a person who is replicated through a malfunctioning digital duplicate) or them alienating their target audience (e.g., in the case when the deployment of a duplicate is viewed by a certain group of stakeholders as inappropriate or offensive). These risks are amplified by the inherent unpredictability of generative AI, which means there is always a risk that a digital duplicate will produce unintended or problematic responses. This partial lack of control connects to ongoing debates in AI ethics regarding 'responsibility gaps' (Matthias 2004; de Sio and Mecacci 2021) and the 'credit-blame asymmetry' (Nyholm, 2024).

The concept of responsibility gaps refers to situations when it is difficult to assign responsibility for harm caused by autonomous systems. In the case of survivors' duplicates, however, the inverse is also true: while a museum curator may not have full control over a duplicate's behavior, they may still be held accountable for the harm it causes. The credit-blame asymmetry highlights another challenge: when AI functions well, its autonomy obscures the contributions of designers and deployers, making it seem as if no one deserves credit for AI's success. While this may not be harmful per se, the lack of recognition can frustrate those involved and lead to a diminished sense of responsibility or an increased willingness to cut corners.

Regarding mitigation strategies, some of the harms discussed in this section may be resolved over time. As AI technology matures and becomes more widespread, the best practices regarding design decisions will likely emerge, forming a set of principles that will be used to identify appropriate guardrails and content moderation practices. However, certain challenges are likely to remain unresolved and potentially contentious, such as balancing the originality of a digital duplicate's outputs with the need to stay faithful to the original testimony or determining the appropriate ways to use this technology in the case of perpetrators (or bystanders). As a result, the design and deployment of AI-enhanced digital duplicates will likely remain controversial for upcoming years and will place a significant burden on those involved due to potential harm.

## Conclusions

In this paper, we examined the emerging phenomenon of digital duplicates and the ethical considerations that arise as this technology is adopted in specific domains. Using the MVPP framework proposed by Danaher and Nyholm (2024), we investigated whether and under which conditions the use of digital duplicates is permissible for enhancing Holocaust survivor testimonies and identified key risks that must be addressed from design and socio-technical perspectives to mitigate potential harms arising from such uses. While doing so, we critically examined the five conditions for duplicates' permissibility and discussed how some of these conditions (e.g., consent and positive value) might need to be problematized when applied to specific scenarios in sensitive domains such as genocide memorialization and education.

Our analysis demonstrates that the deployment of even generative AI-enhanced duplicates of Holocaust survivors meets several permissibility criteria, in particular regarding the need for authentic presence, positive value, and transparency. However, we also demonstrate that for certain conditions, particularly regarding consent and risk-harm mitigation, substantial effort is required to address potential risks and identify fitting mitigation strategies. In the case of consent, the core challenge is related to the inherent potential of the technology: because generative AI can make digital duplicates more flexible than their current rule-based versions, it is difficult to determine what exactly individuals will be consenting to and how to identify whether a specific use is non-consensual (especially in the case of duplicates which can be engaged with virtually on a one-to-one basis). As for risk-harm mitigation, the complexity of meeting this condition is related to the different groups of actors affected by possible harms arising from the use of digital duplicates in the context of Holocaust testimonies. Consequently, multiple mitigation strategies must be applied to minimize these risks, thus amplifying the costs of developing and deploying digital duplicates.

We suggest that our case study offers a valuable lens for evaluating the normative implications of applying digital duplicates in a sensitive context that is also particularly likely to be affected by the current advancements of AI. As the last Holocaust survivors pass away, institutions dealing with Holocaust remembrance and education must confront the challenge of ensuring that their testimonies remain relevant and accessible for future generations. With basic forms of digital duplicates already being introduced, we expect the adoption of more advanced generative AI-based duplicates in this domain to be particularly likely. Under these circumstances, we suggest that it is paramount to proactively identify the risks of adopting the new technology in advance instead of tackling many complex questions associated with these risks post hoc. At the same time, it does not mean that experimenting with digital duplicates should not happen until all these questions are answered: without concrete prototypes and empirical use cases, it is impossible to develop best practices to mitigate risks, and it is important that Holocaust heritage practitioners and educators take the initiative due to them being better equipped to recognize normative challenges of adopting technology for this specific domain.

We began this paper with reference to the Golem and the Dybbuk, two figures that symbolize the fear of losing agency and control. As we saw throughout the paper, these themes repeatedly resurfaced in our analysis of the ethical concerns regarding digital duplication. As is often the case with ethical inquiry, our discussion does not offer a definitive answer regarding the permissibility of digital duplicates in this use case. Nevertheless, the fears invoked by these two figures remain central: will the digital duplicate act as intended, or become an unpredictable and ungovernable virtual Golem? Will the survivor's agency and will power shape the duplicate's outputs, or will other stakeholders come to inhabit and animate the system, transforming it into a digital Dybbuk? Much will depend on how the various mitigation strategies we proposed are implemented at both technical and socio-technical levels. However, we acknowledge that our discussion has not been exhaustive and that new challenges are likely to emerge over time. That being said, we believe that this analysis is one of the necessary steps for assessing not only the use of digital duplicates in the context of Holocaust testimonies, but in demonstrating the type of analysis that is called for in other memory-related use cases.